PoS

Proceedings of Science

# High-energy gamma-ray studying with GAMMA-400

N.P. Topchiev[a1], A.M. Galper[a, b], V. Bonvicini[c], I.V. Arkhangelskaja[b],
A.I. Arkhangelskiy[b], A.V. Bakaldin[a, d], S.G. Bobkov[d], O.D. Dalkarov[a], A.E. Egorov[a],
Yu.V. Gusakov[a], B.I. Hnatyk[e], V.V. Kadilin[b], M.D. Kheymits[b], V.E. Korepanov[f],
A.A. Leonov[a, b], V.V. Mikhailov[b], A.A. Moiseev[g], I.V. Moskalenko[h], P.Yu. Naumov[b],
P. Picozza[i], M.F. Runtso[b], O.V. Serdin[d], R. Sparvoli[i], P. Spillantini[k], Yu.I. Stozhkov[a],
S.I. Suchkov[a], A.A. Taraskin[b], Yu.T. Yurkin[b], and V.G. Zverev[a]

[a] *Lebedev Physical Institute, RU-119991 Moscow, Russia*
[b] *National Research Nuclear University MEPhI, RU-115409 Moscow, Russia*
[c] *Istituto Nazionale di Fisica Nucleare, Sezione di Trieste, I-34149 Trieste, Italy*
[d] *Scientific Research Institute for System Analysis, RU-117218 Moscow, Russia*
[e] *Taras Shevchenko National University, Kyiv, 01601 Ukraine*
[f] *Lviv Center of Institute of Space Research, Lviv, 79060 Ukraine*
[g] *NASA Goddard Space Flight Center and CRESST/University of Maryland, Greenbelt, Maryland 20771, USA*
[h] *Hansen Experimental Physics Laboratory and Kavli Institute for Particle Astrophysics and Cosmology, Stanford University, Stanford, CA 94305, USA*
[i] *Istituto Nazionale di Fisica Nucleare, Sezione di Rome "Tor Vergata", I-00133 Rome, Italy*
[j] *Istituto Nazionale di Fisica Nucleare, Sezione di Florence, I-50019 Sesto Fiorentino, Florence, Italy*

Extraterrestrial gamma-ray astronomy is now a source of new knowledge in the fields of astrophysics, cosmic-ray physics, and the nature of dark matter. The next absolutely necessary step in the development of extraterrestrial high-energy gamma-ray astronomy is the improvement of the physical and technical characteristics of gamma-ray telescopes, especially the angular and energy resolutions. Such a new generation telescope will be GAMMA-400. GAMMA-400, currently developing gamma-ray telescope, together with X-ray telescope will precisely and detailed observe in the energy range of ~20 MeV to ~1000 GeV and 3-30 keV the Galactic plane, especially, Galactic Center, Fermi Bubbles, Crab, Cygnus, etc. The GAMMA-400 will operate in the highly elliptic orbit continuously for a long time with the unprecedented angular (~0.01° at $E_\gamma$ = 100 GeV) and energy (~1% at $E_\gamma$ = 100 GeV) resolutions better than the Fermi-LAT, as well as ground gamma-ray telescopes, by a factor of 5-10. GAMMA-400 will permit to resolve gamma rays from annihilation or decay of dark matter particles, identify many discrete sources (many of which are variable), to clarify the structure of extended sources, to specify the data on the diffuse emission.

[1]Speaker, e-mail: *tnp51@yandex.ru*

## 1 CURRENT GAMMA-RAY STUDY CHALLENGES

### 1.1 Analysis of gamma-ray results according to the Fermi-LAT and ground-based facility data

Since 2008 Fermi-LAT is operating in a near-Earth orbit in the scanning mode and surveying full sky every three hours. Up to now, three catalogs of gamma-ray sources have been published based on the Fermi-LAT observational results: 1FGL [1] and 2FGL [2] for the energy range from 100 MeV to 100 GeV, 3FGL [3] for the energy range from 100 MeV to 300 GeV. Moreover, three catalogs of high-energy gamma-ray sources were published: 1FHL for the energy above 10 GeV [4], 2FHL for the energy range of 50 GeV – 2 TeV [5], and 3FHL for the energy range of 10 GeV – 2 TeV [6].

Figure 1 [7] shows the percentage of the different types of ~3030 gamma-ray sources according to the 3FGL. However, 33% of gamma-ray sources are unidentified and there no the data in the energy range of 20-100 MeV. From [3], it is seen that during four years of the Fermi-LAT operation the real exposition time of the source observations is only ~12% or 1/8 of total operation time and the scanning mode hardly permit to Fermi-LAT observe hour and day variability of sources.

Based on results of gamma-ray observations at energies above 100 GeV by ground-based facilities VERITAS [8], MAGIC [9], H.E.S.S. [10] and others, the TeVCat catalog (http://tevcat.uchicago.edu/) of discrete gamma-ray sources was created, which contains only about 180 sources. Figure 2 shows the composition of Galactic discrete gamma-ray sources recorded by H.E.S.S. (https://www.mpi-hd.mpg.de/hfm/HESS/pages/home/som/2016/01/). It is seen that 47 from 77 sources are not firmly identified.

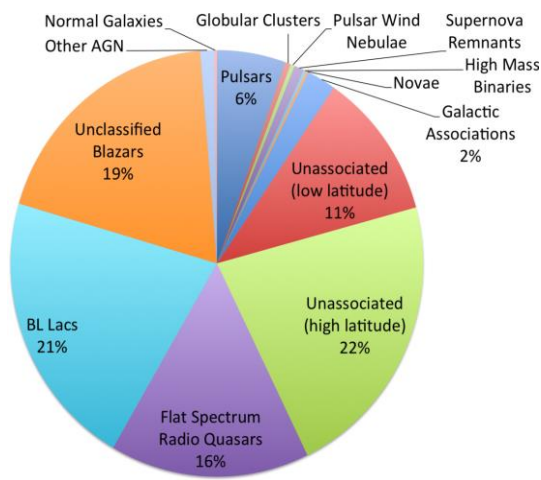
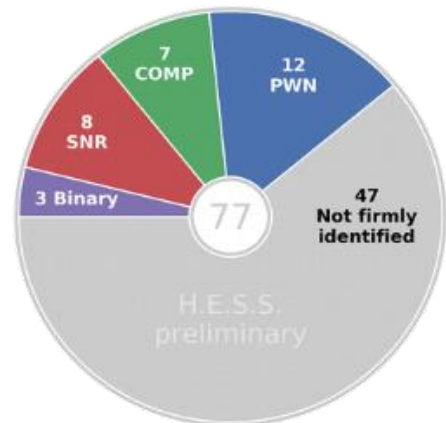

**Fig. 1**. The percentage of the different types of gamma-ray sources [7] according to the 3FGL.

**Fig. 2.** Composition of Galactic discrete gamma-ray sources recorded by H.E.S.S. (https://www.mpi-hd.mpg.de/hfm/HESS/pages/home/som/2016/01/).

It is important to note that the observational data from Fermi-LAT and ground-based facilities were obtained for the energy ranges, which overlap poorly for many gamma-ray sources. Sometimes they do not overlap at all. Hence, the frontier range around 100 GeV is still very interesting for investigations. In addition, the angular resolution of Fermi-LAT, existing ground-based telescopes, and even planned CTA [11] in the region of around 10-300 GeV is only ~0.1°. Therefore, a much better angular resolution is required in order to identify many gamma-ray sources.

**1.2 Indirect searches of dark matter**

Another very interesting and important goal in the studies of gamma-ray sky is indirect searches of dark matter (DM). In general, an exact physical nature of DM is a top puzzle in the modern astrophysics. There are many candidates on the DM role being proposed. However, WIMPs with mass between several GeV and several TeV are still considered as the most probable candidate [12]. WIMPs can annihilate or decay with the production of gamma rays. This emission can have both continuous energy spectrum or monoenergetic lines. This depends on which annihilation channel realizes in the nature. The continuous spectrum would come in the case of annihilation into particle pairs like

$$\chi\chi \to b\bar{b}, \tau^+\tau^-, W^+W^-, \mu^+\mu^-, q\bar{q}, ZZ$$

(Fig. 3, left [13]) or others and gamma-ray lines would be produced in the case of direct annihilation into photons $\chi\chi \to \gamma\gamma, \gamma Z, \gamma H$ (Fig. 3, right [13]).

To resolve gamma-ray lines from background it is necessary to have a high-energy resolution. Figure 4 shows expected energy spectrum for the annihilation of 300-GeV WIMP producing gamma rays ($\gamma\gamma$, $\gamma Z$, and $\gamma H$ lines), which can be resolved from background by various telescopes with the energy resolutions of 10%, 5%, and 0.5% [14]. Note that the energy resolution of Fermi-LAT and ground-based facilities is only 10-15% at the energy of 10-300 GeV. Thus, as seen from Fig. 4, the future telescopes need to have 1-2% energy resolution.

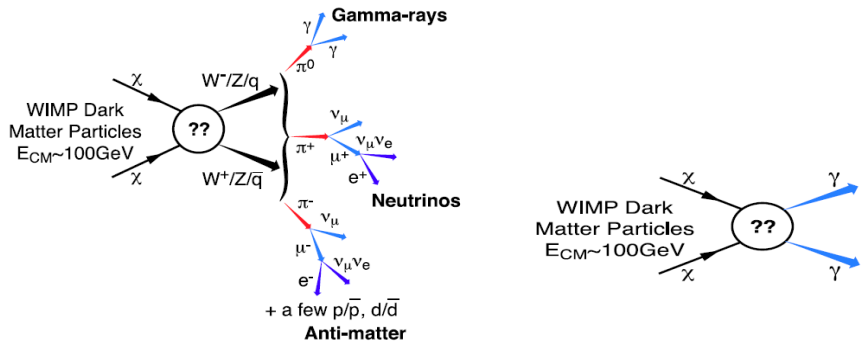

**Fig. 3.** Possible channels of WIMP annihilation $\chi\chi \to b\bar{b}, \tau^+\tau^-, W^+W^-, \mu^+\mu^-, q\bar{q}, ZZ$ with the gamma-ray production [13].

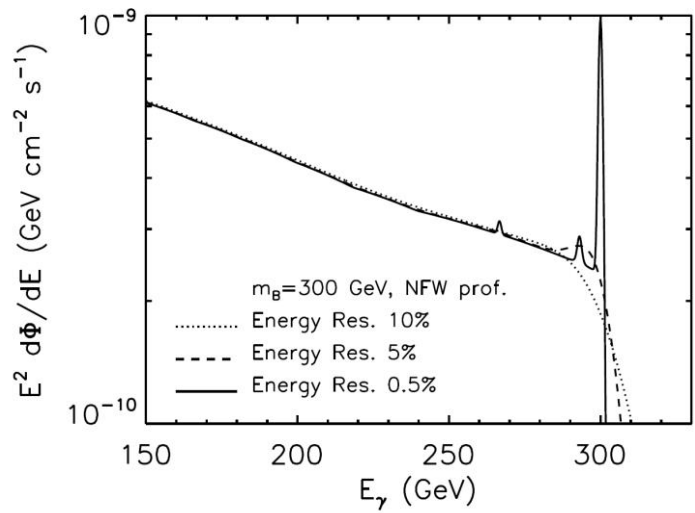

**Fig. 4.** Expected energy spectrum for the annihilation of 300-GeV WIMP producing gamma rays ($\gamma\gamma$, $\gamma Z$, and $\gamma H$ lines), which can be resolved from background by various gamma-ray telescopes with the energy resolutions of 10%, 5%, and 0.5% [14].

## 2 THE GAMMA-400 GAMMA-RAY TELESCOPE

Thus, to resolve unidentified gamma-ray sources and search for the potential gamma-ray lines from DM we need a gamma-ray telescope with the angular resolution of several hundredth degrees and the energy resolution of few percent for the energy of ~100 GeV. Such a new generation telescope will be GAMMA-400, which will be installed onboard the Russian space observatory [15-20].

The GAMMA-400 main scientific goals are: dark matter searching by means of gamma-ray astronomy; precise detailed observations of Galactic plane, especially, Galactic Center, Fermi Bubbles, Crab, Vela, Cygnus, Geminga, Sun, and other regions, extended and point gamma-ray sources, diffuse gamma rays with unprecedented angular (~0.01° at $E_\gamma >$ 100 GeV) and energy (~1% at $E_\gamma >$ 100 GeV) resolutions.

### 2.1 The GAMMA-400 physical scheme and performance

The physical scheme of the GAMMA-400 gamma-ray telescope is shown in Fig. 5. The GAMMA-400 with upgraded converter-tracker can investigate gamma rays from ~20 MeV to ~1000 GeV with the field of view (FoV) of ±45°.

GAMMA-400 consists of plastic scintillation anticoincidence top and lateral detectors (ACtop and AClat), converter-tracker (C), plastic scintillation detectors (S1 and S2) for the time-of-flight system (ToF), calorimeter (CC), plastic scintillation detector (S3).

The anticoincidence detectors surrounding the converter-tracker are used to distinguish gamma rays from significantly larger number of charged particles (e.g., in the region of 10-100 GeV, the flux ratios for gamma rays to electrons and protons are ~ $1:10^2:10^5$.

All scintillation detectors consist from two independent 1-cm layers. The time-of-flight system, where detectors S1 and S2 are separated by approximately 500 mm, determines the top-down direction of arriving particles. The additional scintillation detector S3 improve hadron and electromagnetic shower separation.

The converter-tracker consists of 13 layers of double (x, y) silicon strip coordinate detectors (pitch of 0.08 mm). The first seven layers are interleaved with tungsten conversion foils with 0.1 $X_0$, next four layers with tungsten conversion foils with 0.025 $X_0$ (where $X_0$ is the radiation length). and final two layers have no tungsten. Using the four 0.025 $X_0$ layers allows us to measure gamma rays down to approximately 20 MeV and the gamma-ray polarization. In this case, the gamma-ray trigger for the energy range of 20-100 MeV and 100 MeV – 1000 GeV is the same: $\overline{AC} \times S1 \times S2$. The total converter-tracker thickness is about 1 $X_0$. The converter-tracker information is used to precisely determine the direction of each incident particle.

The two-part calorimeter measures particle energy. The imaging calorimeter CC1 consists of 2 layers of double (x, y) silicon strip coordinate detectors (pitch of 0.08 mm) interleaved with planes from CsI(Tl) crystals, and the electromagnetic calorimeter CC2 consists of CsI(Tl) crystals. The thickness of CC1 and CC2 is 2 $X_0$ and 20 $X_0$, respectively. The total calorimeter thickness is 22 $X_0$ or 1.0 $\lambda_0$ (where $\lambda_0$ is nuclear interaction length). Using a deep calorimeter allows us to extend the energy range up to several TeV for gamma rays and to reach an energy resolution about 1% above 100 GeV.

Figures 6 and 7 show the dependences of GAMMA-400 angular and energy resolutions for the energy range from ~20 MeV to ~100 MeV, for the case, when gamma rays convert in the four 0.025 $X_0$ layers. In this case, the energy release from converter-tracker, S1, and S2 is used.

Figures 8 and 9 show the comparison of angular and energy resolutions for GAMMA-400, Fermi-LAT, H.E.S.S., HAWC, and CTA.

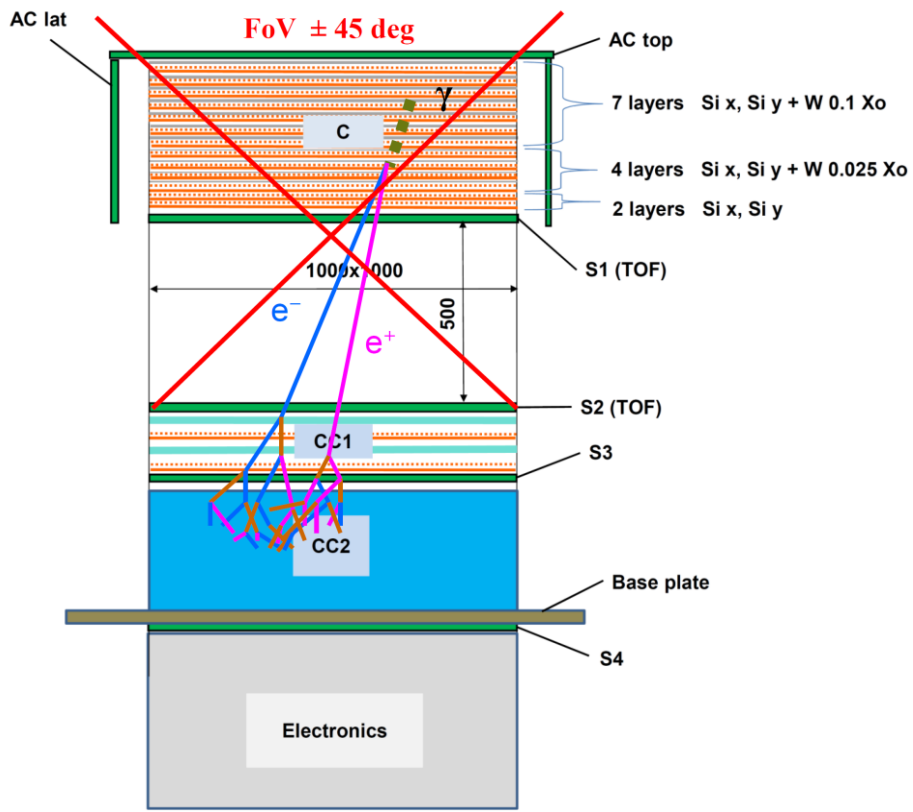

**Fig. 5.** The GAMMA-400 physical scheme.

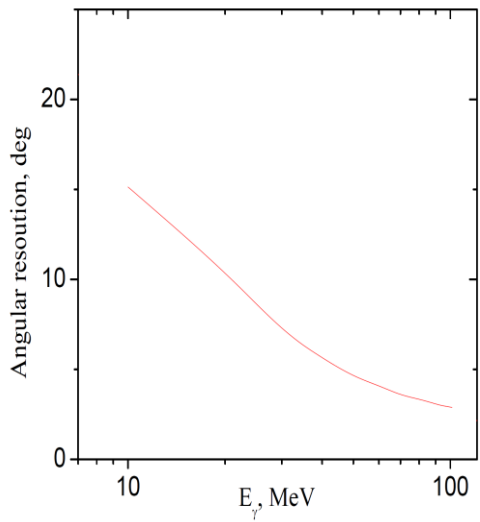

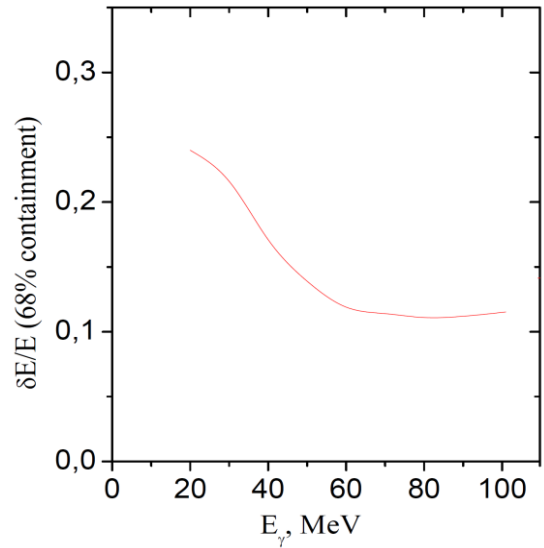

**Fig. 6.** Dependence of the GAMMA-400 angular resolution for the energy range from ~20 MeV to ~100 MeV, for the case, when gamma rays convert in the four 0.025 $X_0$ layers.

**Fig. 7.** Dependence of the GAMMA-400 energy resolution for the energy range from ~20 MeV to ~100 MeV, for the case, when gamma rays convert in the four 0.025 $X_0$ layers.

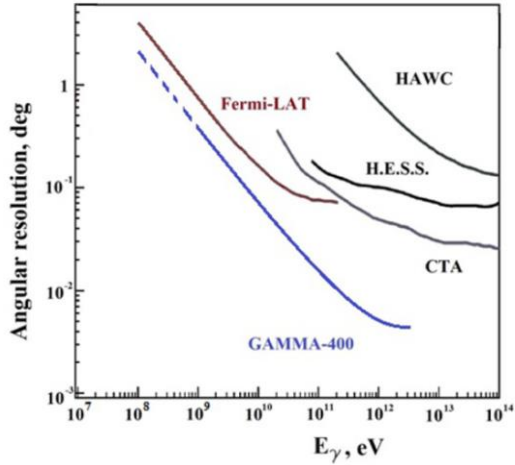
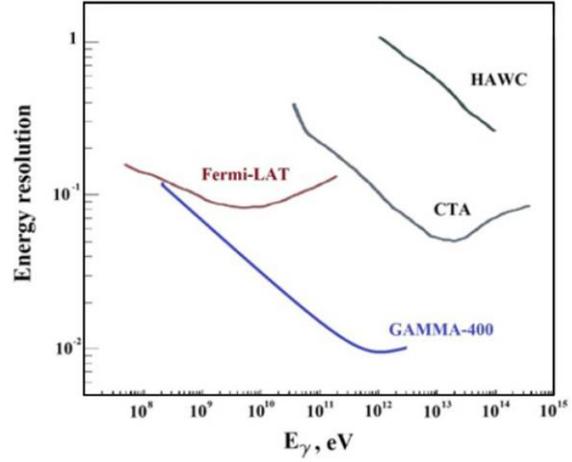

**Fig. 8.** Comparison of angular resolutions for GAMMA-400, Fermi-LAT, H.E.S.S., HAWC, and CTA

**Fig. 9.** Comparison of energy resolutions for GAMMA-400, Fermi-LAT, HAWC, and CTA

GAMMA-400 has numerous advantages in comparison with the Fermi-LAT (Table 1):

• highly elliptical orbit (without the Earth's occultation and away from the radiation belts) allows us to observe with the full aperture of ±45° different gamma-ray sources continuously over a long period of time with the exposition greater by a factor of 8 than for Fermi-LAT operating in the sky-survey mode;

• thanks to a smaller pitch (by a factor of 3) and analog readout in the coordinate silicon strip detectors, GAMMA-400 has an excellent angular resolution above ~20 MeV;

• due to the deep (~22 $X_0$) calorimeter, GAMMA-400 has an excellent energy resolution and can more reliably to detect gamma rays up to several TeV for vertically incident events;

• owing to the better gamma-ray separation from cosmic rays (in contrast to Fermi-LAT, the presence of a special trigger with event timing, time-of-flight system, two-layer scintillation detectors), GAMMA-400 is significantly well equipped to separate gamma rays from the background of cosmic rays and backscattering events.

GAMMA-400 will have also the better angular and energy resolutions in the energy region 10-1000 GeV in comparison with current and future space- and ground-based instruments: VERITAS [8], MAGIC [9], H.E.S.S. [10], CTA [11], and HAWC [21] (Figs. 8 and 9) and it allows us to fill the gap at the energy of ~100 GeV between the space- and ground-based instruments.

GAMMA-400 will study continuously over a long period of time different regions of the Galactic plane, for example, the Galactic center, Fermi Bubbles, Crab, etc. with FoV of ±45. In particular, using the gamma-ray fluxes obtained by Fermi-LAT, we can expect that GAMMA-400, when observing the Galactic center with an aperture of ±45° during 1 year will detect: 57400 photons for $E\gamma > 10$ GeV; 5240 photons for $E\gamma > 50$ GeV; 1280 photons for $E\gamma > 100$ GeV; 535 photons for $E\gamma > 200$ GeV.

The main targets to search for gamma rays from dark matter are:

**The Milky Way.** The center of Milky Way is, apparently, the best potential source of dark matter emission possessing the largest J-factor [12]. Moreover, recently, an anomalous excess of gamma-ray emission in the GeV energy range was revealed near the Galactic center (the region of about one degree) [22], which can be well described by dark matter with a mass of several tens of GeV and annihilation cross section of about standard thermal $10^{26}$ cm$^3$/s. However, this observed excess can have another interpretation - the presence of a population of millisecond pulsars [23]. Therefore, the new GAMMA-400 observational data can help solve this problem.

**Milky Way satellites** have been considered for a long time as the strongest sources of constraints for dark matter, because they have sufficiently large J-factors and at the same time have considerably less gamma-ray background in comparison with the Galactic center.

**Other objects.** Other potentially interesting objects are other galaxies and their clusters, where dark matter may be present and can emit gamma rays. GAMMA-400 with the highest energy resolution of 1% will have a unique sensitivity for detecting dark matter.

Table 1

|  | **Fermi-LAT** | **GAMMA-400** |
|---|---|---|
| Orbit | Circular, 565 km | Highly elliptical, 500-300000 km (without the Earth's occultation) |
| Operation mode | Sky survey (3 hours) | Point observation (up to 100 days) |
| Source exposition | 1/8 | 1 |
| Energy range | ~100 MeV - ~300 GeV | ~20 MeV - ~1000 GeV |
| Effective area ($E_\gamma > 1$ GeV) | ~5000 cm$^2$ (front) | ~4000 cm$^2$ |
| Coordinate detectors - readout | Si strips (pitch 0.23 mm) digital | Si strips (pitch 0.08 mm) analog |
| Angular resolution | ~3° ($E_\gamma$ = 100 MeV) ~0.2° ($E_\gamma$ = 10 GeV) ~0.1° ($E_\gamma$ > 100 GeV) | ~2° ($E_\gamma$ = 100 MeV) ~0.1° ($E_\gamma$ = 10 GeV) ~0.01° ($E_\gamma$ > 100 GeV) |
| Calorimeter - thickness | CsI(Tl) ~8.5 $X_0$ | CsI(Tl) + Si ~22 $X_0$ |
| Energy resolution | ~18% ($E_\gamma$ = 100 MeV) ~10% ($E_\gamma$ = 10 GeV) ~10% ($E_\gamma$ > 100 GeV) | ~10% ($E_\gamma$ = 100 MeV) ~3% ($E_\gamma$ = 10 GeV) ~1% ($E_\gamma$ > 100 GeV) |
| Mass, kg | 2800 | 4100 |
| Telemetry downlink volume, Gbytes/day | 15 | 100 |

### 2.2 The GAMMA-400 space observatory

At the space observatory, along with the GAMMA-400 gamma-ray telescope, an X-ray telescope will be installed. Simultaneous observations in the X-ray and gamma-ray ranges of the Galactic plane, especially, Galactic center, Fermi bubbles, Crab, etc. will greatly improve our understanding of the processes taking place in the astrophysical objects.

The GAMMA-400 space observatory will be installed onboard of the Navigator space platform, which is designed and manufactured by the Lavochkin Association.

Using the Navigator space platform gives the GAMMA-400 experiment a highly unique opportunity for the near future gamma-ray, X-ray, and cosmic-ray science, since it allows us to install a scientific payload (mass of 4500 kg, power consumption of 2000 W, and telemetry downlink of 100 GB/day, with lifetime more than 7 years), which will provide GAMMA-400 with the means to significantly contribute as the next generation instrument for gamma-ray, X-ray astronomy and cosmic-ray physics.

The GAMMA-400 experiment will be initially launched into a highly elliptical orbit (with an apogee of 300,000 km and a perigee of 500 km, with an inclination of 51.4°), with 7 days orbital period. Under the action of gravitational disturbances of the Sun, Moon, and the Earth after ~6 months the orbit will transform to about an approximately circular one with a radius of ~200 000 km and will not suffer from the Earth's occultation and shielding by the radiation belts. A great advantage of such an orbit is the fact that the full sky coverage will always be available for gamma-ray astronomy, since the Earth will not cover a significant fraction of the sky, as is usually the case for low-Earth orbit. Therefore, the GAMMA-400 source pointing strategy will hence be properly defined to maximize the physics outcome of the experiment. The launch of the GAMMA-400 space observatory is scheduled for the middle of the 2020s.